\documentstyle[12pt]{article}\textheight
230mm\textwidth 165mm 
\newcommand{\bi}{\bibitem}
\newcommand{\be}{\begin{eqnarray}}
\newcommand{\ee}{\end{eqnarray}}
\newcommand{\nn}{\nonumber}
\catcode`\@=11
\def\lsim{\mathrel{\mathpalette\@versim<}}
\def\gsim{\mathrel{\mathpalette\@versim>}}
\def\@versim#1#2{\vcenter{\offinterlineskip
\ialign{$\m@th#1\hfil##\hfil$\crcr#2\crcr\sim\crcr } }}
\catcode`\@=12

\begin{document}
\pagestyle{empty}

\noindent
\hspace*{10.7cm} \vspace{-3mm}  HIP-1998-56/TH\\
\hspace*{10.7cm} \vspace{-3mm}  HUB-EP-98/50\\
\hspace*{10.7cm} \vspace{-3mm}  KANAZAWA-98-12\\
\hspace*{10.7cm} \vspace{-3mm}  MPI-PhT/98-65\\

\noindent
\hspace*{10.7cm} August 1998

\begin{center}
{\Large\bf  INFRARED BEHAVIOUR  \\OF\\
SOFTLY BROKEN SQCD AND ITS DUAL}
\end{center}

\begin{center}
{\sc Andreas Karch}$\ ^{(1)}$, 
{\sc Tatsuo Kobayashi}$\ ^{(2),\dag}$, \\
{\sc Jisuke Kubo}$\ ^{(3),*}$ and
{\sc George Zoupanos}$\ ^{(1,4),**}$  
\end{center}
\begin{center}
{\em $\ ^{(1)}$ 
Institut f\" ur Physik, Humboldt-Universit\" at zu Berlin, \vspace{-2mm}
D-10115 Berlin, Germany}\\
{\em $\ ^{(2)}$ 
Department of Physics,  High Energy Physics Division, 
University of Helsinki \vspace{-2mm}\\ and \vspace{-2mm}\\
Helsinki Institute of Physics, 
FIN-00014 Helsinki, Finland} \vspace{-2mm}\\
{\em $\ ^{(3)}$ 
Max-Planck-Institut f\"ur Physik,
 Werner-Heisenberg-Institut \vspace{-2mm}\\
D-80805 Munich, Germany}\vspace{-2mm}\\
{\em $\ ^{(4)}$
Physics Department, Nat. Technical University, GR-157 80 
\vspace{-3mm} Zografou,
Athens, Greece.}
\end{center}

\vspace{0.5cm}
\begin{center}
{\sc\large Abstract}
\end{center}

\noindent
Applying the recently obtained  results on
the renormalization \vspace{-3mm} of soft 
supersymmetry-breaking parameters,
we investigate \vspace{-3mm}
 the infrared behaviour  of the softly broken
supersymmetric QCD as well as \vspace{-3mm} its dual theory
in the conformal window.
Under general assumptions on $\beta$-functions, it is shown 
 that \vspace{-3mm}
the soft supersymmetry-breaking parameters
asymptotically vanish in the infrared limit \vspace{-3mm}  so that
superconformal symmetry in softly broken  supersymmetric 
QCD and  in its dual theory
revives at the infrared fixed point, \vspace{-3mm} provided
the soft scalar masses satisfy
certain renormalization group invariant relations. \vspace{-3mm} 
If these relations are not satisfied, 
there exist marginal
operators  in both theories that lead to the breaking of \vspace{-3mm}
supersymmetry and also  colour symmetry.

\vspace{1cm}
\footnoterule
\vspace{0.1cm}
\noindent
$^{\dag}$Partially supported  by the Academy of Finland \vspace{-3mm} 
(no. 37599).\\
\noindent
$^{*}$On leave from:
Institute for Theoretical Physics, 
Kanazawa \vspace{-2mm} University,  \vspace{-2mm}
Kanazawa 920-1192,  \vspace{-2mm}Japan\\
\noindent
$ ^{**}$
Partially supported  by the E.C. projects, \vspace{-3mm}
FMBI-CT96-1212 and  ERBFMRXCT960090,
the Greek projects, PENED95/1170; 1981.

\newpage
\pagestyle{plain}
%\section{Introduction}
Hunting renormalization group (RG) invariant 
relations among independent parameters in 
a given phenomenological model,
especially supersymmetric one,
has proved in recent years to be successful in
predicting, free otherwise, parameters of the model
 \cite{finite1}--\cite{sum01}.
These are ranging from the top quark mass,
the spectrum of the superpartners to the lightest Higgs
mass in 
$N=1$ supersymmetric finite and non-finite Gauge-Yukawa
unified models  \cite{finite1}--\cite{sum01}.
The basic idea is based on the principle of coupling reduction 
 proposed in ref. \cite{zim1}.
Most of the practical applications of the principle, 
however, have been restricted
to lower orders in perturbation theory,
because the   $\beta$-functions are mostly known only in
lower orders.
Parallel to this development,
interesting progresses has been made on the
renormalization properties of 
the soft supersymmetry-breaking (SSB) parameters 
\cite{yamada1}--\cite{avdeev1}. 
In ref. \cite{yamada1} the spurion formalism in
a softly-broken supersymmetric gauge theory 
has been used to derive a certain set of rules to relate
 the SSB parameters with those of 
the corresponding supersymmetric theory.
Later it has been shown \cite{jack3,avdeev1}
that these rules can be realized
by differential operators acting on
the anomalous dimensions
$\gamma_i$  and the gauge coupling
$\beta$-function $\beta_g$. 
As a result, the $\beta$-functions of the SSB parameters
can be obtained from $\gamma_i$  and
$\beta_g$.
A consequence is that  it has become possible to obtain
a closed form for RG invariant relations among
the SSB parameters \cite{kazakov1}--\cite{jack5}.
Another consequence  is that the asymptotic 
behaviour of the SSB parameters
in the ultraviolet (UV) as well as in infrared (IR) limit can be studied
by using $\gamma_i$ and $\beta_g$ only.
If  $\gamma_i$ and $\beta_g$ are exact, the results on the
SSB parameters derived from them are exact, too.

The purpose of the present paper is to examine on the light of the above
mentioned new development the asymptotic behaviour of 
softly broken supersymmetric QCD (SQCD)  and its 
dual theory  in the IR
limit, especially in the regime for which
according to 
Seiberg's conjecture \cite{seiberg} there would exist 
a non-trivial IR fixed point if supersymmetry was exact.  
The existence of a stable IR fixed point in the space
of the dimensionless couplings (gauge and Yukawa couplings)
of a supersymmetric Yang-Mills theory implies that
the $\beta$-functions 
and their derivatives of these couplings should have
certain properties in the IR limit.
We will use these properties to derive
the asymptotic behaviour of the SSB parameters, and
 will be assuming that
the above mentioned perturbative technique to
obtain their RG functions is applicable to
 the problem we are concerned with. Moreover, we assume that
the kinetic term in the dual theory takes the canonical form.
We will find that
supersymmetry in SQCD
 and also  in its dual
theory restores itself at the IR fixed point, provided
the soft scalar masses satisfy
the RG invariant relations given by (\ref{cond1}) for SQCD, and
(\ref{mq}) and (\ref{cond2}) for its dual theory, respectively.
If these relations are not satisfied, 
there exist marginal
operators which will break
supersymmetry and local colour symmetry in the IR limit.

There exists a rather extensive  literature  
\cite{soft-duality}--\cite{arkani1} on duality of softly
broken $N=1$ SQCD.  Our work is
 related to  those of refs. \cite{cheng1,arkani1}, 
in which the asymptotic behaviour of the SSB parameters has
been also considered.
But their results are obtained mainly in the free magnetic theory.
At the end of our paper we will briefly comment on the relation
to their result.

%\section{Recent results on the renormalization of the SSB parameters}
Following the notation of
 ref. \cite{jack4}, we start with a generic discussion and
consider an $N=1$
supersymmetric gauge theory with the superpontential
\be
W(\Phi) &= &\frac{1}{6} Y^{ijk} \Phi_i \Phi_j \Phi_k + 
\frac{1}{2} \mu^{ij} \Phi_i\Phi_j ~,
\ee
and introduce the SSB part $L_{SSB}$
 \cite{yamada1}:
\be
L(\Phi,W) &=& - \left( ~\int d^2\theta\eta (  \frac{1}{6} 
 h^{ijk} \Phi_i \Phi_j \Phi_k +  \frac{1}{2}  b^{ij} \Phi_i \Phi_j 
+  \frac{1}{2}  MW_A^\alpha W_{A\alpha} )+
\mbox{h.c.}~\right)\nn\\
& &-\int d^4\theta\tilde{\eta} \eta \overline{\Phi^j}                   
(m^2)^i_j(e^{2gV})_i^k \Phi_k~,
\ee
where $\eta = \theta^2$, 
$\tilde{\eta} = \tilde{\theta}^2$ are the external
spurion superfields and $\theta$, $\tilde{\theta}$ 
are the usual grasmannian
parameters, and $M$ is the gaugino mass.
The $\beta$-functions of the $M, h$ and $m^2$
parameters  can be computed from  \cite{jack3,avdeev1}:
\be
\beta_M &=& 2{\cal O}\left({\beta_g\over g}\right)~,
\label{betaM}\\
\beta_h^{ijk}&=&\gamma^i{}_lh^{ljk}+\gamma^j{}_lh^{ilk}
+\gamma^k{}_lh^{ijl}-2\gamma_1^i{}_lY^{ljk}
-2\gamma_1^j{}_lY^{ilk}-2\gamma_1^k{}_lY^{ijl}~,
\label{betah}\\
(\beta_{m^2})^i{}_j &=&\left[ \Delta 
+ X \frac{\partial}{\partial g}\right]\gamma^i{}_j~,
\label{betam2}\\
{\cal O} &=&\left(Mg^2{\partial\over{\partial g^2}}
-h^{lmn}{\partial
\over{\partial Y^{lmn}}}\right)~,
\label{diffo}\\
\Delta &=& 2{\cal O}{\cal O}^* +2|M|^2 g^2{\partial
\over{\partial g^2}} +\tilde{Y}_{lmn}
{\partial\over{\partial
Y_{lmn}}} +\tilde{Y}^{lmn}{\partial\over{\partial Y^{lmn}}}~,
\label{delta}\ee
where $(\gamma_1)^i{}_j={\cal O}\gamma^i{}_j$, 
$Y_{lmn} = (Y^{lmn})^*$, and 
$\tilde{Y}^{ijk}=
(m^2)^i{}_lY^{ljk}+(m^2)^j{}_lY^{ilk}+(m^2)^k{}_lY^{ijl}$.
Note that $X$ in eq. (\ref{betam2})
is explicitly known only in the lowest order \cite{jack6}.
However, there exists an indirect method (which
is based on a RG invariance argument) to
fix the exact form of $X$ \cite{jack4,kkz,jack5},
as we will see below.
We do not consider the $B$ parameters
in the following discussions, because they do not enter into the
$\beta$-functions of the other quantities 
and moreover they are absent in SQCD as well as in its dual theory.

Jack {\em et al.} \cite{jack4},
generalizing the idea of Kazakov \cite{kazakov1}
who has treated a finite theory,
have found 
\footnote{Under the  assumption that
$\gamma^j{}_i = \gamma_i \delta^j{}_i~,~
(m^2)^j{}_i = m^2_i \delta^j{}_i$, and 
$Y^{ijk}(\partial/\partial Y^{ijk})
= Y^{*ijk}(\partial/\partial Y^{*ijk})$
on the space of the RG
functions.} that if
\be
h^{ijk} &=& -M (Y^{ijk})'
\equiv -M \frac{d Y^{ijk}(g)}{d \ln g}~,
\label{h-Y'}\\
  m^2_i &=&
|M|^2 \{(1+\tilde{X}(g)) (g/\beta_g) (\gamma_i (g))+
\frac{1}{2}[(g/\beta_g)\gamma_i (g)]'~\}
\label{m2}
\ee
are satisfied, then the differential operators 
${\cal O}$ and $\Delta$ 
((\ref{diffo}) and (\ref{delta}), respectively) can be
written as \be
{\cal O} &=& \frac{M}{2}
\frac{d }{d \ln g}~,~
\Delta +X \frac{\partial}{\partial g}= |M|^2[~
\frac{1}{2}\frac{d^2}{d (\ln g)^2}
+ (1+\tilde{X}(g))\frac{d}{d \ln g}~]~,
\label{diffo2}
\ee
where
\be
g\tilde{X}(g) &=&\frac{1}{|M|^2} X(g, Y(g), Y^*(g),
h(M,g),h^*(M,g),m^2(|M|^2,g))~.
\label{xtilde}
\ee
It has been 
further shown in ref. \cite{jack4}  that the unknown term
 $\tilde{X}$   (\ref{xtilde})
has to have the form
\be
\tilde{X} &=& \frac{1}{2}(\ln (\beta_g/g))^{\prime}-1~
\label{xtilde1}
\ee
in order 
that the expression (\ref{m2}) is RG invariant. 
Therefore, eq.(\ref{m2}) becomes \cite{jack4},
\be
m^2_i &=&
\frac{1}{2}|M|^2 (g/\beta_g) (\gamma_i (g))'~.
\label{m21}
\ee
In ref \cite{kkz}, with the use of the 
Novikov-Shifman-Vainstein-Zakharov (NSVZ) 
$\beta$-function \cite{novikov1}
for the gauge coupling
\footnote{The factor $2$ of $\gamma_\l$ should
be compare with $1/2$ which appears for
$\beta_g^{\rm NSVZ}$ in ref. \cite{kkz}, where
a typographical error has been made.
There is an indication of the presence of a 
correction term to $\beta_g^{\rm NSVZ}$
in higher order \cite{sibold1}.}
 \be
\beta_g^{\rm NSVZ} &=& 
\frac{g^3}{16\pi^2} 
\left[ \frac{\sum_\l T(R_\l)(1-2\gamma_\l)
-3 C(G)}{ 1-g^2C(G)/8\pi^2}\right]~, 
\label{bnsvz}
\ee 
we have shown that the sum rule 
\footnote{The lower-order sum rules for the soft scalar masses have been
obtained in various theoretical analyses, including
those in superstring models.
See \cite{sum3,sum0,sum01}.}
\be
m^2_i+m^2_j+m^2_k &=&
|M|^2 \{~
\frac{1}{1-g^2 C(G)/(8\pi^2)}\frac{d \ln Y^{ijk}}{d \ln g}
+\frac{1}{2}\frac{d^2 \ln Y^{ijk}}{d (\ln g)^2}~\}\nn\\
& &+\sum_\l
\frac{m^2_\l T(R_\l)}{C(G)-8\pi^2/g^2}
\frac{d \ln Y^{ijk}}{d \ln g}~
\label{sum2}
\ee
is RG invariant  \footnote{Eqs. (\ref{h-Y'}) and (\ref{sum2})
ensure finiteness of the SSB sector in finite theories \cite{kkz}.
In ref. \cite{strassler} duality 
in finite theories has been discussed.},
if the $\tilde{X}$ (\ref{xtilde})
on the subspace defined by 
 $Y=Y(g)$ and eq. $(\ref{h-Y'})$
takes the form \cite{kkz}
\be
\tilde{X} &=&
\frac{-|M|^2 C(G)+\sum_\l m_\l^2 T(R_\l) }{C(G)-8\pi^2/g^2}~.
\label{xtilde2}
\ee
Eq. (\ref{xtilde2}) is consistent with (\ref{xtilde1}) and
the explicit calculations \cite{jack6}, of course.
It has been shown later \cite{jack5} that the restriction
of the constrained subspace
can be removed.
For SQCD (without Yukawa couplings), 
the  $\beta_{m^2}$
in the NSVZ scheme becomes \cite{kkz}
\be
\beta_{m^2_i}^{\rm NSVZ} &=&\left[~
|M|^2 \{~
\frac{1}{1-g^2 C(G)/(8\pi^2)}\frac{d }{d \ln g}
+\frac{1}{2}\frac{d^2 }{d (\ln g)^2}~\}\right.\nn\\
& &\left. +\sum_\l
\frac{m^2_\l T(R_\l)}{C(G)-8\pi^2/g^2}
\frac{d }{d \ln g}~\right]~\gamma_{i}^{\rm NSVZ}~.
\label{bm23}
\ee

%\section{ Supersymmetric QCD and Seiberg's conjectures}

Before we shall come to discuss the
IR behaviour of softly broken 
SQCD and its dual theory,   we would like to 
briefly recall the dynamics of
SQCD and also Seiberg's conjectures \cite{seiberg}.
SQCD is   based on the gauge group $SU(N_c)$,
which contains a gauge superfield $W_{\alpha}$ and quark supermultiplets in
the $(N_c+\overline{N}_c)$ representations of $SU(N_c)$. Let us call the
quark superfields $Q^i_{\alpha}$, $\overline{Q}^{\alpha}_i$, where $i =
1,...N_f$ is a flavour index and $\alpha = 1,...,N_c$ is a colour index. When
these superfields are massless, the theory exhibits the following global
symmetry holding at the quantum level
$SU(N_f)_L \times SU(N_f)_R \times U(1)_B \times U(1)_R$,
where $SU(N_f)_L$ acts on the $Q^i$, $SU(N_f)_R$ acts on the 
$\overline{Q}_i$,
the $U(1)_B$ denotes the baryon number,
 and the $U(1)_R$ denotes the anomaly 
free $R$-symmetry. The superfields transform under the full symmetry group
of the theory 
$G = SU(N_c) \times [~ SU(N_f)_L \times SU(N_f)_R \times U(1)_B
\times U(1)_R ~]$
as follows \newline
$ Q:( N_c ; N_f, 1 ; 1/N_c, (N_f - N_c)/N_f )~,~
 \overline{Q}:( \overline{N}_c ; 1 , \overline{N}_f ; -1/N_c , (N_f -
N_c)/N_f )$, \newline
$W_{\alpha}:( N_c^2 - 1; 1 , 1 ; 0 , 1 )$.

According to  Seiberg's conjecture \cite{seiberg},
SQCD  for $N_f > N_c + 1$ is dual to,  i.e. can be 
described in the IR limit by,
a supersymmetric gauge theory
based on the group $SU(\tilde{N}_c)$ with $\tilde{N}_c = N_f - N_c$,
which couples to the elementary chiral
superfields $q^i$, $\overline{q}_i$, $i = 1,...N_f$, as well as to an
elementary gauge-singlet
superfield $T^{j}_{i}$. The meson superfield $T$ couples to $q$,
$\overline{q}$ via
the  superpotential
\be
\tilde{W} = Y q T \overline{q}~,
 \ee
where $Y$ is a Yukawa coupling.
Without the superpotential the theory would have an additional
global  $U(1)$ symmetry acting on $T$. One can explicitly check that the 
superpotential preserves the anomaly free $R$-symmetry. Therefore, the
newly
constructed theory has the same global symmetry as the original
SQCD. Seiberg \cite{seiberg}
refers to the relation between this theory and the original as
non-Abelian electric-magnetic duality.

Seiberg \cite{seiberg} has further conjectured on 
the existence of an IR fixed
point in the $\beta$-function of SQCD. 
To recall his proposal, consider
the NSVZ $\beta$-function (\ref{bnsvz}) for
SQCD:
 \be
 \beta(g) & =&-{g^3\over 16\pi^2}{3N_c-N_f+2N_f\gamma(g^2)\over   
 1-N_c g^2/8\pi^2}~,~
 \gamma(g^2)=-{g^2\over 16\pi^2}{N_c^2-1\over N_c}+O(g^4)~.
 \ee
 There is a non-trivial zero of the $\beta$-function for
 $N_f=(3-\epsilon)N_c$,
 $N_f,N_c>>1$. At order $\epsilon$, the fixed point $g^2_*$ is given by
 \be   
 N_cg^2_*={8\pi^2\over 3}\epsilon.
\label{g_*}
 \ee
 In fact it was argued in \cite{seiberg} that such a fixed point exists
 in the range ${3\over 2}N_c\leq N_f\leq 3N_c$
\footnote{Such a behaviour
 was already conjectured to hold in ordinary QCD 
in  \cite{BaZa} (see also \cite{oehme1}).
 For a recent discussion on the status of this speculation, see
\cite{KuCaSt}
 and for recent findings in lattice studies,
 see e.g. \cite{Iwas}.}.
 In SQCD the key observation is that the superconformal theory at the
 fixed point has a dual 
 (magnetic) description in terms of a different gauge theory based on
 $SU(N_f-N_c)$.

%\section{ SSB parameters of SQCD in the IR }

Now we come to our main result.
As we will see, the RG invariant relations (\ref{h-Y'})
and (\ref{m21}) are IR attractive,
and they play a crucial role in investigating
the behaviour of the SSB parameters near an IR fixed point.
The sum rule (\ref{sum2}) will be 
used in the dual theory to derive from
the behaviour of the soft scalar masses
near the fixed point (which is obtained
by linearizing the problem) 
a condition which should be satisfied away from the fixed point
in order  to restore supersymmetry 
at the fixed point.
A number of numerical analyses on the IR stability
of the SSB parameters using lower order $\beta$-functions
 have been made previously \cite{infra,sum01}.
Their results are suggestive for our problem.
But we would like to emphasize that
we will discuss the IR stability of the RG invariant relations
(\ref{h-Y'}) and (\ref{m21}) in the conformal window
for which an IR stable fixed point in the space of the gauge coupling
for SQCD, and in the space of the gauge and Yukawa couplings
for the dual theory is supposed to exist.
Our analysis does not relay on the explicit form of
the $\beta$-functions for the SSB parameters.

To begin with, using the formulae
(\ref{betaM})--(\ref{betam2}) and
the RG invariant solutions (\ref{h-Y'}) 
and (\ref{m21}), we show that there always exists
at least a trajectory in the space of the SSB parameters that approaches
the origin if the gauge and Yukawa couplings approach
a non-trivial fixed point.
To this end, we note that 
if eq. (\ref{h-Y'}) is satisfied, then the 
differential operator ${\cal O}$ becomes a total derivative 
operator as we see from eq. (\ref{diffo2}). 
Then eq. (\ref{betaM}) becomes nothing but 
the Hisano-Shifman relation \cite{hisano1}
\be
M &=& M_0  (\beta_g/g)~,
\label{rgi}
\ee
where $M_0 $ is a RG invariant quantity. 
Since $\beta_g \to 0$ as $g \to g_*\neq 0$ by assumption,
we find that $M \to 0$ as $g \to g_*$.
Similarly, eq. (\ref{h-Y'}) implies that
$h^{ijk}=-M_0  (\beta_g/g)(Y^{ijk})^{\prime}
\to 0$. Finally, from eq. (\ref{m21}) we see that
$m_i^2=(1/2) |M_0|^2  (\beta_g/g) (\gamma_i)^{\prime}
\to 0$ as $g \to g_*$.
In what follows, we will investigate carefully whether
the origin of the SSB parameters is a stable IR fixed point.

First we consider
softly broken SQCD
 and examine the IR behaviour of
the gaugino mass $M$ and soft scalar squared masses
$m^2_Q$, $m^{2}_{\overline{Q}}$. 
Since there is no Yukawa coupling
in SQCD, the differential operator (\ref{diffo}) becomes a
total derivative operator, and  we
have 
\begin{eqnarray}
 \beta_M =  M g\frac{d}{d g}(\beta_g/g) 
=M (\beta_g/g)^{\prime}~.
\label{4.2}
\end{eqnarray} 
The conjecture that $g_*$ is a stable
IR fixed point for SQCD implies that
\be
\Gamma_M &\equiv&\frac{d \beta_g }{d g}|_* >  0~,
\label{positivity1}
\ee
where $~~|_*$ means an evaluation at the fixed point.
We may  assume that
\be
|\beta_g| ~~\mbox{and}~~|\frac{d \beta_g }{d g}| & < & \infty~
\label{finiteness1}
\ee
for the range of $g$ we are considering.
Eq. (\ref{4.2}) implies
 the Hisano-Shifman relation \cite{hisano1},
and so if $M_0$ in the r.h. side of (\ref{rgi}) is a non-vanishing
constant, then the gaugino mass $M$ has to vanish at the fixed point,
as we have seen above.
Moreover, one sees from eqs. (\ref{4.2}) and (\ref{positivity1})
that the fixed point $M_*=0$ is a stable one, because
$M \sim e^{ \Gamma_M t} \to 0 ~~\mbox{as}~~t \to -\infty$,
where in the lowest order approximation
 in the $\epsilon$
expansion we have $\Gamma_M =\epsilon^2 /3$.

To discuss the IR behaviour of $m^2_{Q,\overline{Q}}$,
on the same level as $M$, we have to go the NSVZ scheme, and use
the $\beta$-function (\ref{bm23}).
First we would like to show that the RG invariant relation (\ref{m21})
is IR attractive.
To this end, we note that for the $\beta_g^{\rm NSVZ}$  given in
eq. (\ref{bnsvz}), the conditions (\ref{positivity1}) 
and (\ref{finiteness1}) are satisfied if
\be
\Gamma_\gamma \equiv \frac{1}{2}
\frac{d}{d g}(\gamma_Q+\gamma_{\overline{Q}})
|_* &=&\frac{d}{d g}\gamma_Q|_* <  0~\mbox{and}
~N_c-8\pi^2/g^{2}  <   0~
\label{gammagamma}
\ee
are satisfied, where we have used 
$\gamma_Q=\gamma_{\overline{Q}}$.
Then we consider the behaviour of 
 $m^2_{Q}$ and $m^2_{\overline{Q}}$ near the 
RG invariant relation (\ref{m21}),
\be
m^2_{Q,\overline{Q}} &=&m^{2}_{(0)Q,\overline{Q}}
+\delta m^2_{Q,\overline{Q}}~,~
m^{2}_{(0)Q,\overline{Q}}
\equiv \frac{1}{2}|M|^2 (g/\beta^{\rm NSVZ}_{g})
(\gamma_{Q}^{\rm NSVZ})^{\prime}~.
\label{m2Q0}
\ee
Linearizing the evolution equation  near 
the RG invariant relation (\ref{m21}), we find that
\be
\frac{d}{dt }\delta m^2_{Q} &\simeq&
\frac{d}{dt }\delta m^2_{\overline{Q}}~
\simeq \frac{1}{2}\Gamma_{m^2_Q}~(\delta m^2_{Q}
+\delta m^2_{\overline{Q}})~,~
\Gamma_{m^2_Q} \equiv 
\frac{g_*N_f \Gamma_\gamma}{N_c-8\pi^2/g_*^2}~,
\ee
where $\Gamma_\gamma$ is defined in eq. (\ref{gammagamma}).
Since   $\Gamma_{m^2_Q}$
is positive (see eq. (\ref{gammagamma})), we find
\be
\delta m^2_{Q}
-\delta m^2_{\overline{Q}} =\mbox{const.}~,~
\delta m^2_{Q}
+\delta m^2_{\overline{Q}} &\sim& e^{\Gamma_{m^2_Q} t} \to 0
 ~~\mbox{as}~~t \to -\infty~.
\label{irm2}
\ee
(In the lowest order approximation
 in the $\epsilon$
expansion we have $\Gamma_{m^2_Q} =\epsilon^2 /3$.)
Therefore, if the difference $\delta m^2_{Q}
-\delta m^2_{\overline{Q}}$ is non-zero at some point,
then we obtain
\be
\delta m^2_{Q}
&=& -\delta m^2_{\overline{Q}} 
\ee
in the IR limit.
Since, however, $m_{(0),Q,\overline{Q}}^2$ 
(defined in (\ref{m2Q0})) vanish at the fixed point,
we see that
\be
m_{Q}^2 &=& m_{\overline{Q}}^2
\label{cond1}
\ee
should be satisfied in order not to break the colour symmetry in the 
IR limit.
Then from eq. (\ref{irm2}) we may conclude that
\be
m_{Q}^2 &=& m_{\overline{Q}}^2 \sim  e^{\Gamma_{m^2_Q} t} \to 0
 ~~\mbox{as}~~t \to -\infty~.
\label{irm22}
\ee
To conclude, we have shown
that superconformal symmetry revives at the IR fixed point
if $m_{Q}^2 = m_{\overline{Q}}^2$.
Otherwise, the $SU(N_c)$ gauge symmetry and
supersymmetry is broken.

The basic idea for treating the IR behaviour 
of the SSB parameters  of the dual theory
is the same as the  case of SQCD,
where we assume that
the kinetic term in the dual theory takes the canonical form.
A slight difference is that in this case there is a Yukawa
coupling $Y$ in the theory, and hence a trilinear coupling
$h$ in the softly broken case \footnote{The Yukawa
coupling $Y$ is presumably related to the gauge coupling $\tilde{g}$,
because in SQCD there is no such freedom.
In this connection, the reduction of $Y$
in favour of $\tilde{g}$ has been suggested for the dual theory
in ref. \cite{oehme2}.}.

Following Seiberg, we assume that there exists an IR fixed point
in the space of $\tilde{g}$ and $Y$
(the gauge coupling for the dual theory
is denoted by $\tilde{g}$, and the gaugino mass by
$\tilde{M}$). 
The non-triviality of the fixed point of $\beta_Y$ implies 
\be
(\gamma_q
+\gamma_{\overline{q}}+\gamma_T)|_* &=& 0~.
\label{treegamma}
\ee
Further, the stability of the IR fixed point  requires,
among other things, that
\be
\Gamma_{\tilde{M}} &\equiv& 
\frac{d \beta_{\tilde{g}}}{d \tilde{g}}|_*> 0~.
\label{stable1}
\ee
We in addition assume that
\footnote{This is satisfied in the $\tilde{\epsilon}$ expansion.}
\be
2\Gamma_h &\equiv&\frac{\partial \beta_Y}{\partial Y}|_* 
 = \frac{\partial}{\partial Y}[Y (\gamma_q
+\gamma_{\overline{q}}+\gamma_T)]|_* 
=Y_* \frac{\partial}{\partial Y}(\gamma_q
+\gamma_{\overline{q}}+\gamma_T)|_* > 0~.
\label{stable2}
\ee
As in the case of SQCD, we assume that
$\beta_{\tilde{g}}$, $\beta_Y$ together with their derivatives
with respect to $\tilde{g}$ and $Y$ in the space of $\tilde{g}$ 
and $Y$ we are
interested in exist.
For the NSVZ scheme, eq. (\ref{stable1}) means that
\be
\frac{d }{d \tilde{g}}(\gamma_q+\gamma_{\overline{q}})|_*  & < & 0~,
\label{neg1}
\ee
where as in the case of SQCD 
$\tilde{N}_c-8\pi^2/\tilde{g}^{2}  <  0~$ is assumed.

Now we consider the RG invariant relation (\ref{h-Y'})
and show that it is IR attractive.
Defining
\be
h &=&h_0 +\delta h~,~
h_0 = -\tilde{M} Y'=-\tilde{M} 
\tilde{g}\frac{d Y}{d \tilde{g}}
\label{h0}
\ee
and linearizing the evolution equations, we find
\be
\frac{d \tilde{M}}{d t}
&\simeq &\tilde{M}(\beta_{\tilde{g}}
/\tilde{g})^{\prime}-
2\delta h \frac{\partial }{\partial Y}
(\beta_{\tilde{g}}/\tilde{g})~,~
\frac{d \delta h}{d t}
\simeq (1+2Y\frac{\partial }{\partial Y})
[\gamma_q
+\gamma_{\overline{q}}+\gamma_T]~\delta h~.
\ee
So near the fixed point,  $\delta h$ behaves like
$\delta h \sim e^{\Gamma_h t } \to 0 ~~\mbox{as} ~~t \to -\infty$,
where $\Gamma_h$ is defined in eq. (\ref{stable2})
and is a positive number.
Consequently, the gaugino mass behaves like
$\tilde{M} \sim  C_1 e^{\Gamma_{\tilde{M}} t}+C_2 e^{\Gamma_h t}
 \to 0 ~~\mbox{as} ~~t \to -\infty$,
where $C_1$ and $C_2$ are integration constants, and
$\Gamma_{\tilde{M}} (>0)$ are  defined in  eq. (\ref{stable1}).
Therefore, we find that
$\tilde{M}_* = h_*=0$
is a stable fixed point.
In the lowest order approximation
 in the $\tilde{\epsilon}$
expansion we have $\Gamma_{\tilde{M}} 
=\tilde{\epsilon}^2 /3$ and 
$\Gamma_{h} 
=\tilde{\epsilon} /3$, 
where $\tilde{\epsilon}=3-N_f/\tilde{N}_c$.

Next we consider $m^{2}_{q,\overline{q},T}$.
Since near the IR fixed point 
the RG invariant relation 
(\ref{h-Y'}) (or $h_0$ given in eq. (\ref{h0})) is
attractive,  we may use
 $h=h_0$ in the linearization procedure.
We then go to the NSVZ scheme, 
consider a deviation from
 the RG invariant relation (\ref{m21}), and define
\be
m_{i}^{2}=m_{(0)i}^{2}+\delta m_{i}^{2}~,~
m_{(0)i}^{2}\equiv \frac{1}{2}
|\tilde{M}|^2(\tilde{g}/\beta_{\tilde{g}}^{\rm NSVZ})
(\gamma_{i}^{\rm NSVZ})^{\prime}~ ,
\label{mis}
\ee
where $i=q,\overline{q},T$.
We find
\be
\frac{d}{dt }\delta m^2_{q} &\simeq &
\frac{d}{dt }\delta m^2_{\overline{q}}~
\simeq \frac{1}{2}\Gamma_{m^{2}_{q}}~(\delta m^2_{q}
+\delta m^2_{\overline{q}})~,~
\frac{d}{dt }\delta m^2_{T} \simeq 
 \frac{1}{2}\Gamma_{m^{2}_{T}}~(\delta m^2_{q}
+\delta m^2_{\overline{q}})~,
\label{problem1}
\ee
where 
\be
\Gamma_{m^{2}_{q}} &\equiv&
\frac{\tilde{g}_*N_f \Gamma_{\gamma_{q}}}
{\tilde{N}_c-8\pi^2/\tilde{g}_*^2}~,~
\Gamma_{m^{2}_{T}} \equiv
\frac{\tilde{g}_*\Gamma_{\gamma_{T}}}
{\tilde{N}_c-8\pi^2/\tilde{g}_*^2}~,~
\Gamma_{\gamma_{q,T}}\equiv(d \gamma_{q,T}^{\rm NSVZ}/d \tilde{g})|_*~,
\label{gammas3}
\ee
and we have used $\gamma_{q}^{\rm NSVZ}
 =\gamma^{\rm NSVZ}_{\overline{q}}$.
As one can easily find, there are two zero eigenvalues and
one positive one ($=\Gamma_{m^{2}_{q}}$)
in the linearized problem (\ref{problem1}).
One of the two  zero eigenvalues corresponds to the
solution that the difference
$\delta m^2_{q}-\delta m^2_{\overline{q}}$
is constant independent of $t$.
So, as in the case of SQCD, we see that
the dual colour symmetry is broken unless
\be
 m^2_{q}& =&  m^2_{\overline{q}}
\label{mq}
\ee
is satisfied.
The other zero eigenvalue expresses the fact that
$ m^2_{T}$ may contain a piece which is 
constant independent of $t$ in the IR limit.
The presence of the constant part breaks supersymmetry 
at the IR fixed point.
If the relation (\ref{cond1}) in the softly broken 
SQCD is satisfied (so  that supersymmetry is recovered 
at the IR fixed point), we have to demand  
the dual theory, too,  to be
supersymmetric at the IR fixed point.
Therefore, we have the unique solution to
(\ref{problem1}) which preserve supersymmetry at the fixed point:
\newline
$delta m^2_{q} =\delta m^2_{\overline{q}}~,~
\delta m^2_{T}\sim e^{\Gamma_{m^{2}_{q}}t} \to 0
~~\mbox{as}~~t \to -\infty$
with
\be
\frac{\delta m^2_{q}}{\delta m^2_{T}}
&\simeq& \frac{\Gamma_{m^{2}_{q}}}{\Gamma_{m^{2}_{T}}}=
\frac{\Gamma_{\gamma_{q}}}{\Gamma_{\gamma_{T}}}~,
\label{ratio1}
\ee
where $\Gamma$'s are defined in eq. (\ref{gammas3}),
and $\Gamma_{m^{2}_{q}} > 0$
because $\Gamma_{\gamma_q} < 0$ 
if $\gamma_q=\gamma_{\overline{q}}$ (see eq. (\ref{neg1})).
In the lowest order approximation
 in the $\tilde{\epsilon}$
expansion we have $\Gamma_{m^{2}_{q}} =\tilde{\epsilon} /3$.

The ratio $\delta m^2_{q}/\delta m^2_{T}$ being 
constant independent of $t$
suggests the existence of a RG invariant
relation. 
In fact (\ref{ratio1}) is the consequence of the RG invariant
sum rule (\ref{sum2}).
To see this, we insert $m^{2}_{i}~~(i=q,\overline{q},T)$
(\ref{mis}) with $m^{2}_{q}=m^{2}_{\overline{q}}$ into the sum rule
(\ref{sum2}), and find that the sum rule reduces to
\be
2 \delta m^{2}_{q}+\delta m^{2}_{T}
&=& \frac{1}{C(\tilde{G})-8\pi^2/\tilde{g}^2}~
[(\tilde{g}/\beta_{\tilde{g}}^{\rm NSVZ})
(2 N_f \gamma^{\rm NSVZ}_{q}+\gamma^{\rm NSVZ}_{T})]~\delta m^{2}_{q}~.
\label{sum4}
\ee
In the IR limit, the quantity in [ ~~] on
the r.h. side contains an expression $0/0$.
To obtain the correct limit, we  compute
\be
\frac{(d/d \tilde{g})(2 N_f \gamma^{\rm NSVZ}_{q}+\gamma^{\rm NSVZ}_{T})}
{(d/d \tilde{g})(\beta_{\tilde{g}}^{\rm NSVZ}/\tilde{g})}~
\label{limit1}
\ee
at the fixed point. We find that the expression (\ref{limit1})
at the fixed point can be written  as
\be
(C(\tilde{G})-8\pi^2/\tilde{g}^2)~(~2+
\frac{\Gamma_{\gamma_{T}}}{\Gamma_{\gamma_{q}}}~)~,
\ee
implying that the sum rule (\ref{sum4})
exactly becomes (\ref{ratio1}).
Therefore, the soft scalar masses away from the IR fixed point
have to satisfy the sum rule
\be
 m^{2}_{q}+m^2_{\overline{q}}+m^{2}_{T}
&=&
|\tilde{M}|^2 \{~
\frac{1}{1-\tilde{g}^2 C(\tilde{G})/(8\pi^2)}
\frac{d \ln Y}{d \ln \tilde{g}}
+\frac{1}{2}\frac{d^2 \ln Y}{d (\ln \tilde{g})^2}~\}\nn\\
& &+
\frac{ (N_f/2)(m^{2}_{q}+m^2_{\overline{q}})}
{C(\tilde{G})-8\pi^2/\tilde{g}^2}
(\frac{d \ln Y}{d \ln \tilde{g}})~
\label{cond2}
\ee
and also (\ref{mq}) 
so that all the soft scalar masses asymptotically vanish
in the IR limit.
As a result, superconformal symmetry 
in the dual theory, too,  revives at the IR fixed point.
If eq. (\ref{cond1}) for SQCD, and
eqs. (\ref{mq}) and (\ref{cond2}) for the dual theory
are not satisfied, there will be marginal operators that
break supersymmetry as well as
the local gauge symmetries in the IR limit.

As we have mentioned at the beginning, our work is related
to those of refs. \cite{cheng1,arkani1}.
In ref. \cite{cheng1,arkani1}, the relations of the SSB parameters
in the electric and magnetic sides outside of the conformal
window have been derived.
Interestingly, it was found \cite{arkani1}
that the magnetic soft scalar
masses vanish at the boundary between the free magnetic 
and conformal window, which is consistent with our
finding. Note that the sum rule (\ref{cond2})
at the lowest order in $\tilde{g}$ becomes 
$m^{2}_{q}+m^2_{\overline{q}}+m^{2}_{T}=|\tilde{M}|^2$.
It is interesting to compare this with the sum rule
$ m^{2}_{q}+ m^{2}_{\overline{q}}+m^{2}_{T}=0$
obtained outside of the conformal window 
in ref. \cite{cheng1,arkani1},
from which  it has been concluded that 
 the vacuum structure with the soft scalar masses
differs from that without them.
This should be contrasted to our result that in the conformal
window the vacuum structure remains the same as 
long as the relations  (\ref{cond1}) , (\ref{mq}) and (\ref{cond2}) 
are satisfied, supporting the duality hypothesis 
in the presence of the SSB terms in the conformal window.

\vspace{1cm}
We would like to thank 
D. L\" ust and R. Oehme for useful discussions.

\end{document}